\documentstyle[epsfig]{mn}

\newcommand{\Msolar}{\mbox{\,$\rm M_{\odot}$}}        

\newcommand{\kms}{km~s$^{-1}$}
\begin{document}
\title[]{[O III] emission line in narrow-line Seyfert 1 galaxies}
\author[Bian Weihao, Yuan Qirong \& Zhao Yongheng]
{W.Bian$^{1}$, Q. Yuan$^{1}$ and Y.Zhao$^2$\\
$^{1}$Department of Physics and Institute of Theoretical Physics,
Nanjing Normal University, Nanjing
210097, China\\
$^{2}$National Astronomical Observatories, Chinese Academy of
Sciences, Beijing 100012, China\\
}
\maketitle
\begin{abstract}
Three sets of two-component profiles are used to simultaneously
model the [O III]$\lambda\lambda$4959, 5007 and H$\beta$ lines for
the Fe II-subtracted spectra of 149 narrow-line Seyfert 1 galaxies
(NLSls) from the Sloan Digital Sky Survey (SDSS). Using the
linewidth of the narrow/core component of [O III]$\lambda$5007 to
trace the stellar velocity dispersion instead of using the total
linewidth of [O III]$\lambda$5007, we found that the SDSS NLSls
are still deviated from the $M_{bh}-\sigma$ relation found in the
nearby inactive galaxies, suggesting that the linewidth of the [O
III] narrow/core component is likely not a good tracer of bulge
velocity dispersion in NLSls, since some other studies indicate
that NLS1s, like other active galactic nuclei (AGNs), should
follow the $M_{bh}-\sigma$ relation. If we assume that the [O
III]5007/H$\beta_{n}$ line ratio emitted in narrow line region
(NLR) ranges from one to ten, 63 objects are found to satisfied
with this criterion and their H$\beta$ broad components should be
used to calculate their virial black hole masses. These 63 objects
are statically consistent with the $M_{bh}-\sigma_{[O III]}$
relation. With the Chandra observation of some SDSS NLSls, for one
object in these 63 objects, J143030.22-001115.1, we found that it
can't be classified as a genuine NLS1. Its narrow component of
H$\beta$ is coming from NLRs. This is consistent with its very
flat hard X-ray spectrum found by researchers.

\end{abstract}

\section{INTRODUCTION}

There is a strong relation between the central black hole mass and
the bulge velocity dispersion (the $M_{bh} - \sigma$ relation) for
inactive nearby galaxies (Gebhardt et al. 2000a; Ferrarese \&
Merritt 2000; Tremaine et al. 2002), suggesting that the formation
and evolution of host galaxies and their active nuclei are
intimately related. For active galactic nuclei (AGNs),
reverberation mapping method and then the empirical
size-luminosity relation are usually used to measure the black
hole mass instead of the gas and stellar dynamics used in nearby
galaxies (Peterson 1993; Kaspi et al. 2000). Broad-line AGNs
follow this relation (Gebhardt et al. 2000b; Ferrarese 2001;
Nelson 2001; Shield et al. 2003; Boroson 2003; Bonning et al.
2005; Greene \& Ho 2005c). Many theoretical models are presented
to explain the established $M_{bh} - \sigma$ relation, considering
the regulation of the bulge growth by the feedback from the
accretion around the black hole (e.g. Silk \& Rees 1998; King
2003; Hopkins et al. 2005).

However, for an interesting subclass of AGNs, narrow-line Seyfert
1 galaxies (NLSls), their locus in $M_{bh} - \sigma$ plane is
still a question to debate. NLS1s are defined with the following
characteristics. H$\beta$ full width at half-maximum (FWHM) less
than $2000$ \kms; strong optical Fe II multiplets; line ratio of
[O~III]$\lambda$5007 to H$\beta$ less than 3 (Osterbrock \& Pogge
1985; Goodrich 1989); steep, soft X-ray excess (Puchnarewicz et
al. 1992; Boller, Brandt \& Fink 1996) and rapid soft/hard X-ray
variability (Leighly 1999). We also note that the soft X-ray
photon indices of some NLS1s observed by Chandra are found to be
not too steep compared with that normally observed in NLS1s
(Williams et al. 2004). NLS1s are believed to have less massive
black holes with higher Eddington ratios, suggesting that NLS1s
might be in the early stage of AGN evolution (Grupe 1996; Mathur
2000; Bian \& Zhao 2003).

NLSls seemed not to follow the $M_{bh} - \sigma$ relation if [O
III] linewidth is used to trace the bulge velocity dispersion
$\sigma$ (Mathur, Kuraszkiewicz \& Czerney 2001; Bian \& Zhao
2004a; Grupe \& Mathur 2004). NLS1s locus in the $M_{bh} - \sigma$
plane possibly depend on some parameters, such as their accretion
ratios (Mathur \& Grupe 2005). Greene \& Ho (2004) presented a
sample of 19 AGNs with low-mass black holes from Sloan Digital Sky
Survey (SDSS) Data Release One (DR1). These 19 AGNs can be
classified as NLS1s because of their H$\alpha$ FWHM less than
2000\kms. Barth, Greene \& Ho (2005) measured $\sigma$ in these 19
NLS1s. They found these NLS1s follow the $M_{bh} - \sigma$
relation. The linewidth of [O III] indeed typically overestimates
$\sigma$ comparing to the direct measurement of $\sigma$. Botte et
al. (2005) also reaches this result.

As we known, the [O III] profile is usually bluewards asymmetric,
i.e. with more flux on the short-wavelength side of line than on
the long-wavelength side (Peterson 1997). And the strong Fe II
multiples would blend the [O III] and H$\beta$ lines in NLS1s.
Multi-component profile and Fe II template are needed to model the
[O III] lines in NLS1s. Greene \& Ho (2005a) recently suggested
that the core of [O III] after removing its asymmetric blue wing
can trace $\sigma$ in narrow line (type 2) galaxies. Is it true
for NLS1s?

We used the largest published sample of 150 NLSls to investigate
this problem (Williams, Pogge \& Mathur 2002). Their spectra have
been analyzed using multi-component model to investigate the [O
III] blueshift in NLS1s, and we found seven "blue outliers" (Bian,
Yuan \& Zhao 2005a). In this paper, We want to investigate whether
NLS1s follow the $M_{bh} - \sigma$ relation when we used the
narrow/core component of [O III] line to trace $\sigma$. In Sec.
2, we briefly introduce the data and the analysis. Our results and
discussion are given in Sec. 3. A conclusion is presented in the
final section. All of the cosmological calculations in this paper
assume $H_{0}=75 \rm {~km ~s^ {-1}~Mpc^{-1}}$, $\Omega_{M}=0.3$,
$\Omega_{\Lambda} = 0.7$.

\section{DATA AND ANALYSIS}
There are many samples of NLS1s: (1) an optically selected sample
of 46 NLS1s with extremely steep soft X-ray spectra observed with
ROSAT (Boller et al.1996); (2) a compiled  sample of 64 NLS1s
(Veron-Cetty, Verron \& Cloncalves 2001); (3) a sample of 150
NLS1s found within SDSS Early Data Release (EDR) (Williams et al.
2002); (4) 50 NLS1s from a complete sample of 110 soft X-ray
selected AGNs (Grupe et al. 2004); (5) 19 AGNs with low-mass black
holes from SDSS DR1 presented by Greene \& Ho (2004). Here we used
the 150 SDSS NLS1s sample because it is the largest published
NLS1s sample. Because of the lack of the [O III] line, SDSS
J153243.67-004342.5 is ignored in our analysis.

Considering strong Fe II multiples and the asymmetry of [O
III]/H$\beta$ lines, we reduced their SDSS spectra by the
multicomponent fitting task SPECFIT (Kriss 1994) in the IRAF-STS
package. The components are(1) the Galactic interstellar reddening
curve; (2) Fe II template; (3) power-law continuum; (4) three sets
of two-gaussian profiles for [O III]$\lambda\lambda$4959, 5007 and
H$\beta$ lines. For the doublet [O III]$\lambda \lambda$4959,5007,
We take the same linewidth for each component, and fix the flux
ratio of [O III]$\lambda$4959 to [O III]$\lambda$5007 to be 1:3.
We didn't consider the starlight contribution because of no
obvious stellar lines (Gu et al. 2005). For more details, please
refer to Bian, Yuan \& Zhao (2005a).

\section{RESULTS AND DISCUSSION}

\subsection{Distribution of $\Delta
\lambda=\lambda_{broad}-\lambda_{narrow}$ for the H$\beta$ and [O
III] lines}
\begin{figure}
\begin{center}
\epsfig{figure=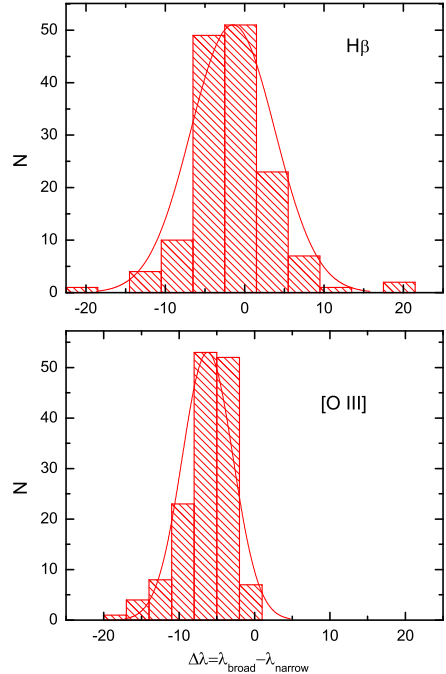,width=7cm}

\caption{The distribution of $\Delta
\lambda=\lambda_{broad}-\lambda_{narrow}$ for the H$\beta$ and [O
III] lines.}
\end{center}
\end{figure}

As we mentioned above, the [O III] profile is usually bluewards
asymmetric. We calculated the blueshift of the broad component
relative to the narrow component ($\Delta
\lambda=\lambda_{broad}-\lambda_{narrow}$) for the [O III] and
H$\beta$ lines. In Fig.1, we showed the distribution of $\Delta
\lambda$ for the H$\beta$ and [O III] lines. It is obvious that
the [O III] profiles tend to be bluewards while the H$\beta$
profiles tend to be bluewards or redwards.

\subsection{$M_{\rm H\beta} - FWHM^{n}([O III])$}

\begin{figure}
\begin{center}
\epsfig{figure=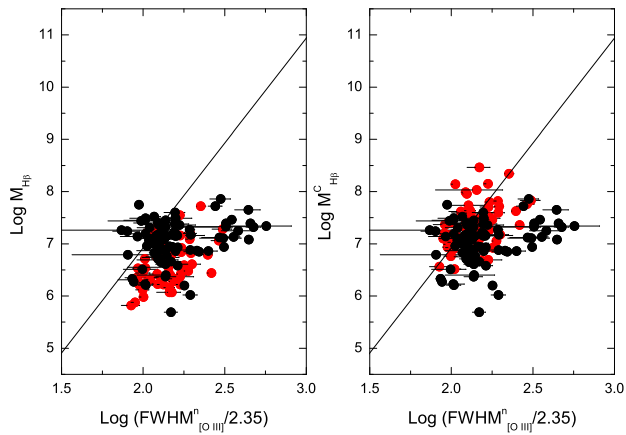,width=9.5cm}

\caption{The relation between the black hole masses and the width
of the narrow component of [O III] line. Left-hand side: the black
hole masses are calculated using FWHM of H$\beta$ derived from
one-Gaussian profile fitting, which are adopted from Bian \& Zhao
(2004a). The red circle denotes the object with [O
III]$\lambda$5007/H$\beta_{n}$ line ratio larger than 1. The solid
line shows the $M_{bh}-\sigma$ relation defined by Tremaine et al.
(2002). Right-hand side: the black hole masses are corrected for
the objects with [O III]$\lambda$5007/H$\beta_{n}$ line ratio
larger than 1, which are calculated using FWHM of the broad
component of H$\beta$ line. }
\end{center}
\end{figure}

In our two-component model, we found that for some objects the
optical Fe II multiples seriously blend with the lines of H$\beta$
and [O III]$\lambda\lambda$4959, 5007. In some cases, Fe II
multiples showed almost the same magnitude of flux to that of [O
III] line (see fig. 3 in Bian, Yuan \& Zhao 2005a).

Bian \& Zhao (2004a) directly measure the [O III]$\lambda$5007
linewidth [hereafter $FWHM^{one}([O III])$] using one-Gaussian
profile model. The spectrum resolution R is about 1800, which is
equivalent to 166\kms. The typical error of $FWHM^{one}([O III])$
is about 10 per cent. For the width of narrow component of [O
III]$\lambda$5007 line [hereafter $FWHM^{n}([O III])$] from
two-component model, the typical error is about 10 per cent.

In the left panel of Fig. 2, we plotted the central black hole
masses ($M_{H\beta}$) versus $FWHM^{n}([O III])$. The mass versus
$FWHM^{one}([O III])$ is showed in fig. 1 in Bian \& Zhao (2004a).
We adopted the same range of x and y axes in order that we can
compare the results with that in fig. 1 in Bian \& Zhao (2004a).
The masses are calculated from the H$\beta$ linewidth using
one-Gaussian fit, which are form Bian \& Zhao (2004a).

Grupe \& Mathur (2004) also plot the $M_{bh} - \sigma_{[O III]}$
relation for a complete sample of 75 soft X-ray¨Cselected AGNs: 43
broad-line AGNs and 32 NLS1s. They found that the locus of NLS1s
obviously deviates from the $M_{bh}-\sigma_{[O III]}$ relation
defined by Tremaine et al. (2002). Considering the blue asymmetry
of the [O III] profile, they remeasured the width of the [O III]
line as two times half-width at half-maximum of the red part of
emission line and found the deviation indeed exists. In the
left-hand panel of Fig. 2, we still find this result, which is
consistent with Grupe \& Mathur (2004).

\subsection{Mass correction}

\begin{table}
\begin{center}
\begin{tabular}{lllllll}
\hline \hline
Type & number & log($M_{H\beta}/M_{[O III]}$) & SD\\
\hline
Total    &     149   & -0.77$\pm$0.06     &    0.70 \\
Subsample A & 63    & -0.90$\pm$ 0.06     &    0.51 \\
Subsample A & 63    & -0.13$\pm$ 0.07$^{\star}$     &    0.53 \\
Subsample B & 86    & -0.67$\pm$0.09      &    0.79 \\
Subsample C & 54    &-0.09$\pm$0.07      &    0.53 \\
\hline \hline
\end{tabular}
\caption{The distributions of log($M_{H\beta}/M_{[O III]}$) for
149 SDSS NLSls. Sub-sample A: the 63 objects with [O
III]$\lambda$5007/H$\beta_{n}$ line ratio larger than one.
Subsample B: the rest 86 NLSls. Subsample C: Subsample A excluding
9 objects with the linewidth of the H$\beta$ broad component less
than 2000 \kms. $^{\star}$: the corrected black hole mass from the
FWHM of the H$\beta$ broad component.}
\end{center}
\end{table}

When we used the linewidth of H$\beta$ or H$\alpha$ to trace the
virial velocity around black hole, we should subtract the
contribution from NLRs. The template built from [O III] or [S II]
is used to model narrow H$\alpha$ and H$\beta$ (Grupe et al. 1998;
Grupe, Thomas \& Leighly 1999; Greene \& Ho, 2005a,b). For seven
NLS1s, Rodriguez-Ardila et al. (2000) found that the narrow
component of H$\beta$ is about, 50\% of the total line flux and
the [O III] $\lambda$5007/H$\beta_{n}$ ratio emitted in the narrow
line regions (NLRs) varies from 1 to 5, instead of the universally
adopted value of 10. We also found the [O III] is not too weak in
many SDSS NLSls. This is consistent with the results of a sample
of 64 NLSls presented by Veron-Cetty et al. (2001). There are 63
SDSS NLSls with [O III]$\lambda$5007/H$\beta_{n}$ line ratio
larger than one. If we assume the narrow H$\beta$ is emitted from
NLRs for these objects, we should use the linewidth of the
H$\beta$ broad component to calculate the virial black hole
masses, which are showed in the right panel of Fig. 2.

We also calculated the black hole mass, $M_{[O III]}$, using FWHM
of narrow component of [O III] line as the indicator of $\sigma$,
i.e. $\sigma_{[O III]} = FWHM^{n}([O III])/2.35$.
$$M_{[O III]} = 10^{8.13}(\sigma_{[O III]}/(200 km s^{-1}))^{4.02} \Msolar$$

The distributions of log($M_{H\beta}/M_{[O III]}$) for 149 SDSS
NLSls are shown in Table 1, where $M_{H\beta}$ is calculated from
H$\beta$ FWHM using one-Gaussian fitting. These 149 NLSls
statically deviated the $M_{bh}-\sigma_{[O III]}$ relation defined
by Tremaine et al. (2002)(See Fig. 2). Considering the spectrum
resolution, the intrinsic $\sigma$ derived from $FWHM^{n}([O
III])$ may be instrumentally broadened by about 60 \kms (hereafter
$\sigma_{inst}$) (Greene \& Ho 2005a). The values of $\sigma$
derived from $FWHM^{n}([O III])$ for all objects in Fig. 1 are
larger than 60 \kms. To first order, the intrinsic $\sigma$ can be
approximated by $\sigma=(\sigma_{obs}^2-\sigma_{inst}^2)^{1/2}$.
We found that the logarithm value of intrinsic $\sigma$ would be
lowered by 0.08 dex, which is small relative to the deviation in
Fig. 2 (also see Table. 1). Subsample A consists of 63 objects
with [O III]$\lambda$5007/H$\beta_{n}$ line ratio larger than 1.
Subsample B consists of the rest 86 NLSls. If we used the width of
the H$\beta$ broad component to calculated the black hole masses,
we found that these 63 objects in Subsample A follow the $M_{bh} -
\sigma_{[O III]}$ relation. In these 63 objects, we found nine
objects with the linewidth of the H$\beta$ broad component less
than 2000\kms . If we excluded these nine objects from sub-sample
A, i.e. Subsample C, the mean value of log($M_{H\beta}/M_{[O
III]}$) would be smaller, -0.09$\pm$0.07 with a standard deviation
of 0.53. Therefore, it is possible that these 54 objects in
Subsample C are not genuine NLSls. It needs a more careful
H$\beta$ subtraction from NLRs contribution in future.

\subsection{$FWHM^{n}([O III])$ is not a good tracer in NLSls?}

From Fig. 2, we found that the locus of NLS1s obviously deviates
from the $M_{bh}-\sigma_{[O III]}$ relation defined by Tremaine et
al. (2002). If the linewidth of the [O III] narrow/core component
overestimates $\sigma$ and then [OIII] is not a good tracer of
$\sigma$, this suggests the particular environment of NLRs in
NLS1s comparing with other AGNs. On the other hand, if the narrow
[OIII] component does trace $\sigma$, then our results showed that
NLS1s possibly do lie below the $M_{bh}-\sigma$ relation (Grupe \&
Mathur 2005).

However, the values of $\sigma$ for some NLS1s are directly
measured from the CaII/Mg b absorption lines(Filippenko \& Ho
2003; Barth et al. 2004; Botte et al. 2005; Barth et al. 2005),
these NLS1s follow the $M_{bh}-\sigma$ relation in a statistical
sense, where the mass is calculated from H$\beta$ FWHM. Bian \&
Zhao (2004b) also found that the mass from the soft X-ray bump
luminosity is consistent with that from the H$\beta$ FWHM for
NLSls. Therefore there is no underestimate in mass calculation
using H$\beta$ FWHM. This showed that the [O III] line is likely
not a good tracer of bulge velocity dispersion.

Greene \& Ho (2005a) investigated the relation between the
velocity dispersion and the line width of [O III] with the sample
of narrow-line (Type 2) AGNs from SDSS DR2. They found that, after
the asymmetric blue wing is properly removed, the width of the [O
III] core component can be used as a tracer of stellar velocity
dispersion. They also looked for the secondary parameters for the
$\Delta \sigma \equiv log\sigma_{[O III]}-log\sigma$ and found a
correction equation  $\Delta \sigma=0.072 log
L_{bol}/L_{Edd}+0.08$. Considering $L_{bol}/L_{Edd}$ is in the
range of -1 $\sim$ 1 (see fig.1 in Bian Yuan \& Zhao 2005a), the
$\sigma$ correction would be 0.008 $\sim$ 0.152, which is small
relative to the deviation in Fig. 2 (also see Table. 1). Mathur \&
Grupe (2005) also used this correction to derive $\sigma$ from
$\sigma_{[O III]}$. However, they found that NLS1s and BL AGNs are
still significantly different (see their fig. 1). They suggested
some NLS1s with high Eddington ratios deviate from the
$M_{bh}-\sigma_{[O III]}$ relation and reside preferentially in
relatively late type galaxies.

\subsection{J143030.22-001115.1 with flat X-ray spectrum}

Williams et al. (2004) suggested that the soft X-ray photon
indices of some SDSS NLSls observed by Chandra are found not too
steep compared with that normally observed in NLSls. There are two
objects with the photo indices less than one. One is
J125943.59+010255.1 ($\Gamma=0.25^{+0.80}_{-1.01}$ ) and the other
is J143030.22-001115.1 ($\Gamma =0.92\pm 0.64$).  They belong to
our 149 SDSS NLSls. For J125943.59+010255.1, the net 0.5-8 kev
count rate is smallest among their Chandra observations and the
uncertainties of $\Gamma$ is too large. The [O III] is too weak in
the SDSSS spectrum of J1259+0102. For J143030.22-001115.1, its [O
III] line is obvious. [O III]$\lambda$5007/H$\beta_{n}$ line ratio
is 9.236, the largest one in these 149 SDSS NLSls. However, the Fe
II line is very weak. Therefor we think the narrow component of
the H$\beta$ line is coming from NLRs. We used three-components to
model the H$\beta$ line. The widths of two narrow components of
H$\beta$ were forced to be equal with that of [O III]$\lambda$5007
line. And the shifts between the corresponding two-component of
H$\beta$ and [O III] were fixed to be 146\AA. It is found that the
FWHM of the H$\beta$ broad component is 2783\kms. Therefore it is
a broad line AGN, a misclassified NLS1. Some authors also found
some wrong classified NLSls (e.g. Veron- Cetty et al. 2001; Botte
et al. 2004). We should be cautious about using the narrow
component of H$\beta$ to trace the black hole masses. We should
exclude this kind of object in the statistics of NLSls. For more
details, please refer to Bian, Cui \& Chao (2005b). It is still a
question whether soft X-ray photon indices of NLS1s found in SDSS
are not too steep compared with that normally observed in NLSls.
It needs more X-ray observations of SDSS NLS1s. We would
discussion these questions in our future paper.

\section{CONCLUSION}

Three sets of two-component profiles are used to model the [O III]
$\lambda\lambda$4959, 5007 and H$\beta$ lines for 149 SDSS NLSls.
The main conclusions can be summarized as follows.

\begin{itemize}

\item{Using the linewidth of the [O III] narrow/core component, we
found that 149 SDSS NLSls are still deviated from the
$M_{bh}-\sigma_{[O III]}$ relation found in nearby inactive
galaxies. If the linewidth of the [O III] narrow/core component
overestimates $\sigma$ and then [OIII] is not a good tracer of
$\sigma$, this suggests the particular environment of NLRs in
NLS1s comparing with other AGNs. On the other hand, if the narrow
[OIII] component does trace $\sigma$ then our results showed that
NLS1s possibly do lie below the $M_{bh}-\sigma$ relation.}

\item{If the [O III]/H$\beta$ line ratio from NLRs is between one
to ten, we found that the narrow H$\beta$ in 63 objects is from
NLRs. we used the broad H$\beta$ line width to calculate the black
hole masses. And these 63 objects follow the $M_{bh}-\sigma_{[O
III]}$ relation found in nearby inactive galaxies. Excluding nine
objects (FWHM(h$\beta$) less than 2000 \kms) from these 63
objects, the rest of 54 objects seemed not to be genuine NLSls.}

\item{J143030.22-001115.1 can't be classified as a genuine NLSls.
Its narrow component of H$\beta$ is coming from NLRs. After the
H$\beta$ contribution from BLRs removed, H$\beta$ FWHM is
2783\kms. This is consistent with its very flat hard X-ray
spectrum founded by Williams et al. (2004).}

\end{itemize}

\section*{ACKNOWLEDGMENTS}

We thank Luis C. Ho for his very helpful comments. We thank the
anonymous referee for the valuable comments. This work has been
supported by the NSFC (No. 10403005; No. 10473005; No. 10273007)
and NSF from Jiangsu Provincial Education Department (No.
03KJB160060). Funding for the creation and distribution of the
SDSS Archive has been provided by the Alfred P. Sloan Foundation,
the Participating Institutions, NASA, the National Science
Foundation, the US Department of Energy, the Japanese
Monbukagakusho, and the Max Planck Society. The SDSS Web site is
http:// www.sdss.org/. This research has made use of the NASA/IPAC
Extragalactic Database, which is operated by the Jet Propulsion
Laboratory at Caltech, under contract with NASA.

\end{document}